# Ta-based Josephson junctions using insulating ALD TaN tunnel barriers

Ekta Bhatia[1], Jack Lombardi[3], Tuan Vo[1], Michael Senatore[3],
Alexander Madden[3], Soumen Kar[1], Hunter Frost[2], Stephen Olson[1],
Jakub Nalaskowski[1], John Mucci[1], Brian Martinick[1], Ilyssa Wells[1],
Thomas Murray[1], Kevin Musick[1], Corbet S. Johnson[1], Stephen McCoy[3],
Daniel L. Campbell[3], Matthew D. LaHaye[3], and Satyavolu S. Papa Rao[1]

[1]NY Creates, Albany, NY 12203, USA
[2]College of Nanotechnology, Sci. & Eng., University at Albany (SUNY), Albany, NY 12203, USA
[3]Air Force Research Laboratory, Information Directorate, Rome, New York 13441, USA

Corresponding authors: Ekta Bhatia, Satyavolu S. Papa Rao (email: ebhatia@ny-creates.org, spaparao@ny-creates.org).

This work was supported by the Air Force Research Lab (AFRL), Rome, NY, under Grant FA8750-19-1-0031 and Grant FA864921P0773.

Josephson junctions form the core circuit element in superconducting quantum computing circuits, single flux quantum digital logic circuits, and sensing devices such as SQUIDs. Aluminum oxide has typically been used as the tunnel barrier. Its formation by exposure to low oxygen pressures at room temperature for short periods of time makes it susceptible to aging and limits the thermal budget of downstream processes. In this paper, we report the first demonstration of α-Ta/insulating TaN/ α-Ta superconductor/insulator/superconductor (SIS) Josephson junctions fabricated on 300 mm wafers using CMOS-compatible processes. The junctions were fabricated on high-resistivity silicon substrates using standard processes available at 300 mm scale, including 193 nm optical lithography, atomic layer deposition (ALD) of TaN in a cluster tool, and chemical mechanical planarization (CMP) to enable highly planar interfaces. Junction areas ranging from 0.03 µm$^2$ to 9 µm$^2$ with ALD TaN thickness between 2 nm and 7 nm were characterized. A critical current density ($J_c$) of 76 µA/µm$^2$ was observed in junctions using 4 nm ALD TaN in the tunnel barrier. The dependence of $J_c$ on ALD TaN layer thickness is analyzed, and the influence of junction geometry, packaging, and temperature on I-V characteristics is discussed. Junctions were retested after a period of 4 months to quantify junction aging. The potential of this novel material system and a 300 mm superconducting junction process flow to fabricate thermally and environmentally stable junctions is discussed. The vision of a Superconducting Quantum Process Design Kit (SQPDK) for a Multi-Project Wafer (MPW) program to enable rapid development and proliferation of superconducting quantum and digital digital logic systems is presented. This work represents the first step towards establishing such a Quantum Foundry, providing access to high quality qubits and single-flux quantum (SFQ) logic circuits at 300 mm wafer scale.

## I. INTRODUCTION

Josephson junctions (JJs) form the heart of superconducting quantum computing circuits [1], single-flux quantum logic circuits [2], and even the NIST voltage standard [3]. Traditionally, Josephson junctions have been formed by oxidation of Al under tightly controlled conditions at room temperature [4]. SFQ circuits have used Nb as the interconnect due to its high superconducting transition temperature ($T_c$), but with SFQ circuits being considered as control chips [5] for quantum computing chips with operating temperatures well below 4.2 K, other materials like α-phase Ta become viable candidates for such applications. Recently, Ta has proven advantageous for use in superconducting qubits [6-8]. Ta is well known in integrated circuit (IC) fabrication facilities, having been introduced together with copper interconnects in the late 1990's [9]. Hence the development of Ta based JJs using 300 mm wafer fabrication tools appears timely, since it could satisfy a multiplicity of requirements. This work focuses on α-Ta electrodes combined with an insulating ALD TaN tunnel barrier, offering a thermally robust and chemically stable alternative platform compatible with similar manufacturing infrastructure.

It is well-known that Al/AlO$_x$/Al qubits suffer from a low thermal budget, with their characteristics degrading after exposure to temperatures greater than 433 K [10]. The aluminum oxide (AlO$_x$) tunnel barrier is also sensitive to poisoning caused by hydrogen out-diffusion from Nb [11]. Additionally, the AlO$_x$ layer formed by room-temperature oxidation of Al necessarily follows the grains and grain boundaries of the underlying metal, resulting in topographical variation and making reproducibility a long-standing problem [12]. Recent efforts on large-scale superconducting qubit fabrication at 300 mm scale [24] have demonstrated Al/AlO$_x$/Al junctions. Even so, these considerations suggest that replacing AlO$_x$ formed by room temperature oxidation with another tunnel barrier scheme might offer advantages. ALD Al$_2$O$_3$ has been reported as a possible tunnel barrier candidate for Al-based junctions [14], and AlN, deposited by magnetron sputtering has been considered in epitaxial NbN/AlN/NbN junctions [33]. Reactively sputtered TaN has been used as a conductive barrier in Nb/TaN/Nb SNS junctions, with $I_c$ and $R_N$ tunable via nitrogen flow; however, non-monotonic trends and interfacial reactions indicate process-control challenges in that stack. As SNS devices, these junctions exhibit $I_cR_N$ products lower than SIS Nb/AlOx/Nb junctions [15]. In contrast, ALD TaN, widely used in the IC industry as a diffusion barrier [16], offers the promise of forming thermally stable interfaces with Ta and of providing greater resistance to oxidative aging. ALD TaN is also much more resistant to attack [17,18] by various chemistries (dilute hydrofluoric acid, buffered oxide etch, or sulfuric-peroxide mixtures) that could be used to clean the surfaces of chips with fabricated junctions before packaging and testing. When the ALD TaN is deposited on a planarized Ta surface [20], the topography issues described above are completely avoided. However, processes and materials developed for CMOS applications cannot be transplanted wholesale—they must be modified to fit the unique requirements of JJs when used in qubits and SFQ circuits. For example, ALD TaN used in JJs needs to be insulating, while for CMOS applications it is intentionally conductive [16,19]. The present α-Ta/insulating ALD TaN/α-Ta JJ demonstrates tunneling behavior with the expected exponential dependance of $J_c$ on thickness, confirming the suitability of this stack for both SFQ and qubit applications.

After describing the fabrication sequence for α-Ta/ALD TaN/ α-Ta JJs using 300 mm wafer scale tools and processes, this paper describes the results of DC characterization of junctions at both room-temperature and cryogenic temperatures. The paper concludes with a description of ongoing and future work, such as the development of a Superconducting Process Design Kit (SQPDK). Such a PDK can enable the community of researchers engaged in superconducting quantum and SFQ circuits to utilize a Quantum Foundry to fabricate their designs using state-of-the-art materials, processes and tools.

## II. EXPERIMENTAL

This section outlines the junction fabrication process, highlighting where new processes are required for superconducting applications. The process steps where there are differences when Ta-based JJs are used in SFQ circuits versus qubit applications are also pointed out – the flow as developed can be modified for either case. It includes a description of the JJ designs tested, along with an overview of cryogenic measurement tools and the measurement procedures utilized.

### A.   *Josephson Junction Fabrication*

The junction fabrication process starts with patterning of trenches by 193 nm immersion optical lithography followed by reactive ion etch (RIE) of a dielectric stack of 15 nm silicon oxide (formed by wet thermal oxidation of high-resistivity silicon) and 50 nm SiN deposited by rapid thermal chemical vapor deposition (RTCVD) to expose the underlying silicon. This is followed by an anisotropic aqueous etch of the exposed silicon to form 80 nm-deep trenches with sloped sidewalls defined by the [111] silicon planes, as shown in Fig. 1(a). This geometry was selected to enable subsequent Josephson junction integration into qubit devices. The use of such chemie douce processes for qubit fabrication allows for surfaces to be created with no silicon damage. We note that replacing trenches wet-etched into silicon with trenches patterned by RIE in plasma-enhanced chemical vapor deposition (PECVD) silicon oxide can be readily implemented, and would be applicable for junctions used in high-performance SFQ logic circuits.

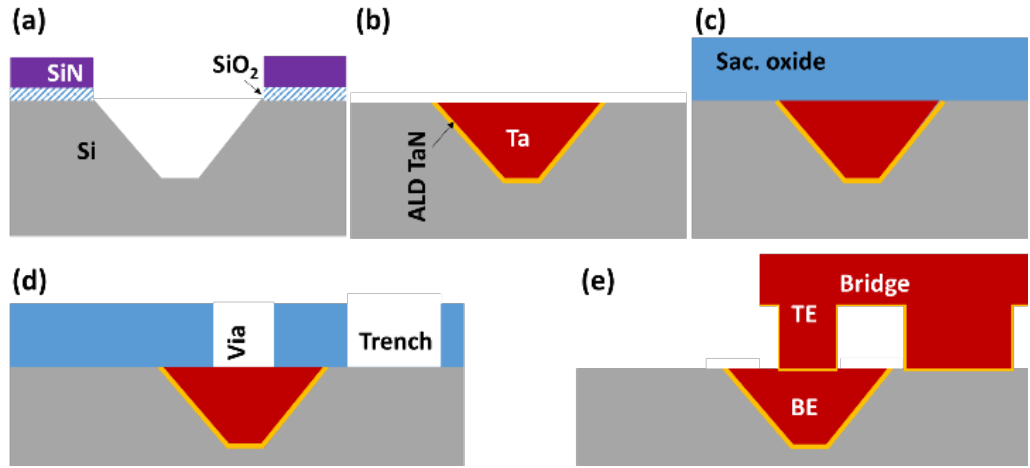

FIGURE 1. JJ fabrication process flow: (a) After trench formation in silicon (b) Post Ta CMP (c) After sacrificial oxide deposition (d) After bottom Ta is exposed to define junction areas (e) Completed Josephson junction (TE indicates top electrode and BE indicates bottom electrode and JJ is the Josephson junction formed at via). A bridge connects the JJ to the rest of the circuit. Both BE and TE are α-Ta.

After removal of SiN with hot phosphoric acid, the wafer is covered with 250 nm of superconducting α-Ta. In order to promote the formation of α-phase Ta, a 3 nm thick TaN film is deposited by atomic layer deposition [20], immediately before the in-situ sputter deposition of Ta in another chamber of the cluster tool. The overburden of Ta (above the plane of the silicon oxide) is then removed using chemical mechanical planarization (CMP), resulting in damascene α-Ta features as shown in Fig. 1(b). CMP is one of the process steps that requires significant work to translate capabilities developed by the CMOS industry into one suitable for superconducting circuits [20].

After CMP, the wafer surface has damascene α-Ta traces in a field of thermal oxide 5 nm to 10 nm thick. For quantum applications, where two-level system (TLS) defects in oxides are detrimental, this oxide can be removed by dilute hydrofluoric acid etch; for SFQ logic circuits, such an oxide removal step is not relevant. The damascene α-Ta traces have widths ranging from 100 nm to 5 μm in circuit-related features, and areas as large as 0.0225 mm$^2$ in process-metrology-related features (which are not 'cheesed', as described in [20]).

This is followed by the deposition of a 100 nm thick PECVD oxide layer. This stage of processing is illustrated by Fig. 1(c). In the case of Josephson junctions that form part of a qubit, the 100 nm PECVD oxide layer is sacrificial, being etched away at the end of device fabrication to eliminate TLS contributions to qubit decoherence, however, it must remain in place for structural integrity in SFQ circuits that typically have over 10 subsequent metallization levels.

Fig. 1(d) shows a schematic cross-section after vias and trenches are formed in the oxide by a sequence of lithography, RIE and wet etch processes. This exposes the underlying Ta at the via bottoms and both Si and Ta at the trench bottoms. Josephson junction fabrication is then completed by tunnel barrier and top electrode deposition, following sputter cleaning of the Ta oxide in the vias. As part of this process flow, junctions were fabricated with ALD TaN tunnel barrier thicknesses ranging from 2 nm to 7 nm. The specifics of ALD TaN tunnel barrier formation for Josephson junctions are covered by a patent disclosure [21]. Josephson junctions formed at the bottom of via as described above have several desirable characteristics and differ from the process flow utilized by other researchers [22]. Since the junction is placed well within the planarized region of the damascene α-Ta forming the bottom electrode, the junction interface is very planar. In addition, the size of the junction is controlled by one lithographic exposure. The final steps consist of top-electrode patterning by 193 nm lithography, RIE, and removal of sacrificial oxide with dilute hydrofluoric acid. Fig. 1(e) shows the schematic of the Josephson junction investigated in this work. In contrast to Josephson junctions patterned by subtractive RIE, this process flow eliminates the risk of metal tendrils at the edge causing shunts across the tunnel barrier.

## B. Josephson Junction Designs

The layout for JJ is shown in Fig. 2(a), with cross-sectional transmission electron micrograph in Fig. 2(b). These junctions use a 'bridge' of α-Ta to connect the junction to the rest of the circuit. The tested junction geometries were varied from 180 nm to 3000 nm and bridge length was varied from 100 nm to 400 nm.

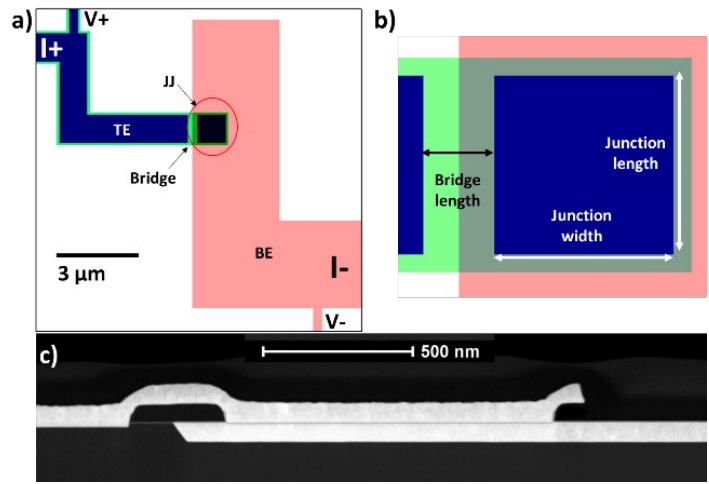

FIGURE 2. (a) Layout, (b) Zoomed-in view showing the bridge, and (c) TEM cross-section of a 1 μm² JJ indicating planarity over 1 um² area.

## C. Cryogenic Measurement Set Up

Cryogenic testing was conducted in an adiabatic demagnetization refrigerator (ADR) at 100 mK, after the chips were diced out, and enclosed in a package that permitted 6 junctions to be measured in each cool-down. When high currents were applied to junctions with thinner barriers (2 nm ALD TaN), the recorded temperature rose, temporarily to as high as 240 mK during testing, but for most conditions, the recorded temperature was between 100 mK and 150 mK. The chips with tested junctions were packaged into ceramic chip carriers using cryogenic varnish as an adhesive and 17 μm Al wirebonds used for electrical connection. The chip carrier was inserted to a socket on a printed circuit board, which provided electrical connection and mechanical mounting to the refrigerator with copper standoffs for test, as shown in Fig. 3. The increase in temperature during I-V characterization is an indication of self-heating. We note that the temperature reported is from the cold stage, with the thermal resistance between the stage and the device under test implying that the actual device temperature is higher.

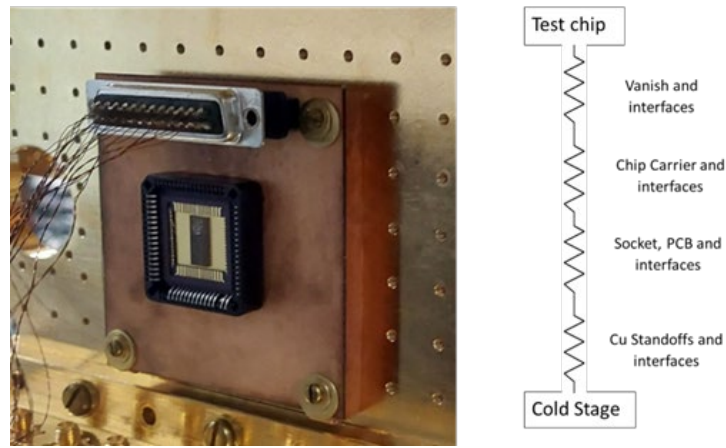

FIGURE 3.    Chip on cold stage, and schematic of thermal resistances to cold stage.

Cryogenic measurements of I-V traces were made by incorporating cryogenic low-pass passive filters (manufactured by QDevil) on each line of a 4-wire measurement. Use of such a cryogenic low-pass filter was found to be critical – without such filtering, important features such as I-V hysteresis were not visible.  A Keithly 2600A source-measure unit was used to perform electrical measurements with python scripts to automate sweeps and collect data.

## III. RESULTS AND DISCUSSION

Characterization of the devices was done at both room temperature and at cryogenic temperatures. Room temperature measurements are faster, enabling considerably more devices to be tested, and help quantify the degree of uniformity of junction characteristics across the 300 mm wafer.

### A. Room Temperature Characterization

Room temperature measurements of junction resistance were made in a 4-wire Kelvin measurement configuration, with a current forced across the junction at a value commensurate with the expected resistance, and voltage was measured with high-impedance contacts to the bottom and top electrodes.

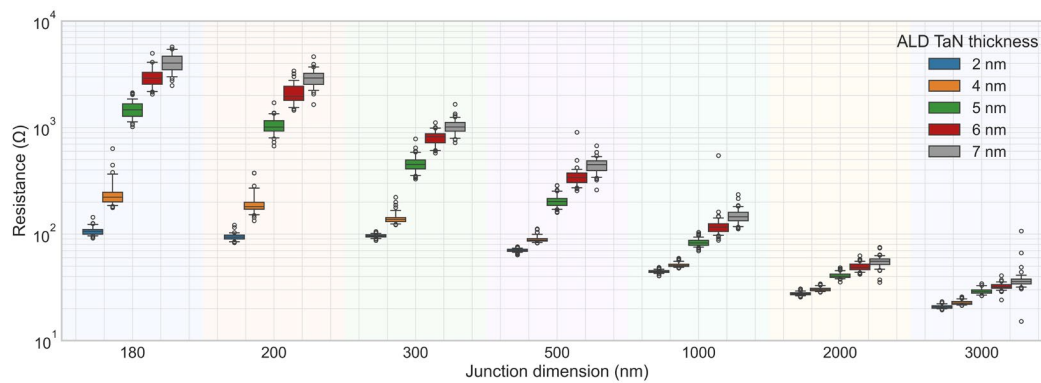

FIGURE 4.  Box-and-whisker plot of room-temperature resistance for square Josephson junctions with different widths. Each color band corresponds to a junction with the nominal lateral dimension (width = length) indicated on the x-axis. Whiskers indicate 5th and 95th percentiles, and open circles indicate outliers beyond these limits. Data are taken from 78 junctions for each width, except for 180 nm junctions, where 52 devices were measured.

The resistance distributions for square junctions with different widths are shown in Fig. 4, using measurements taken at each of 26 dies distributed across wafers with ALD TaN thickness from 2 nm to 7 nm. Each die has 3 copies of each junction width except for the 180 nm wide junctions where 2 copies are present. It should be noted that with the current fabrication process, there is a difference between the actual and the targeted junction lateral dimensions of approximately 40 nm — so a square-shaped junction that is nominally 200 nm wide has an actual length and width of 240 nm. This difference (dimensional bias) is anticipated to be reduced with further process optimization. Fig. 4 shows that resistance scales inversely with junction area and increases exponentially with barrier thickness, consistent with tunneling behavior. The resistance variation increases as the junction size decreases, which can be explained by the relative sensitivity of smaller junctions to a given magnitude of CD variation. This motivates the within-wafer non uniformity (WIWNU) analysis in Fig. 5. The WIWNU of JJ resistance is given by the standard deviation as a percentage of the median resistance, (calculated from measurements of 78 junctions for every width, except for 180 nm, which uses 52 junctions).

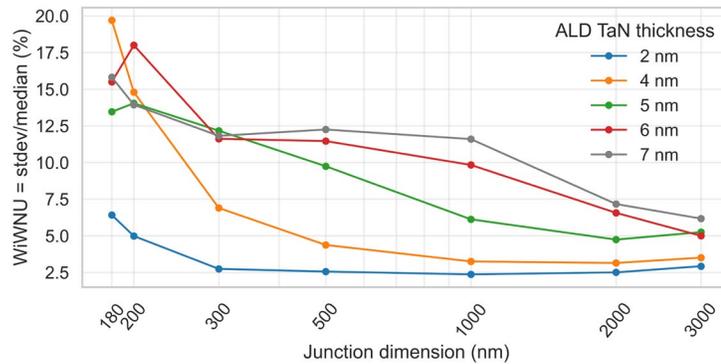

FIGURE 5. Within wafer non-uniformity of room temperature JJ resistance for square junctions with different areas. The labels on x-axis indicate the nominal lateral dimension (width = length) of the square junction.

Sources of resistance variation include junction size (arising from post-RIE critical-dimension variation across the wafer) and lead resistance (due to post-CMP thickness variations) that remain even in this 4-wire Kelvin configuration. The resistance distributions in Fig. 5 indicate that Ta/TaN based junctions show wafer-scale variability comparable to the 8–20% reported for Al/AlOx/Al junctions fabricated using 193 nm lithography and subtractive patterning [24]. A direct comparison is difficult given the order of magnitude difference in room temperature resistivity of the Ta compared with Al. There are differences in fabrication processes and data presentation practices across groups [23,24]. Our work demonstrates that α-Ta/ALD TaN junctions can be fabricated with high yield at 300 mm and offer advantages in terms of thermal stability, conformality, and chemical robustness that are beneficial for 3D integration.

The yield of junctions was > 95%, using ±1.5 times the inner-quartile range as the metric for the defining 'yield' at this preliminary stage of technology development. Yield of JJs remained high irrespective of the bridge length, up to the 0.4 µm length included in JJ structures. No "droop" was observed in 45° tilt-view SEM even at the longest bridge length (1.0 µm) included in bridge-length test structures [25]. Bridge length was therefore not a significant parameter influencing junction characteristics.

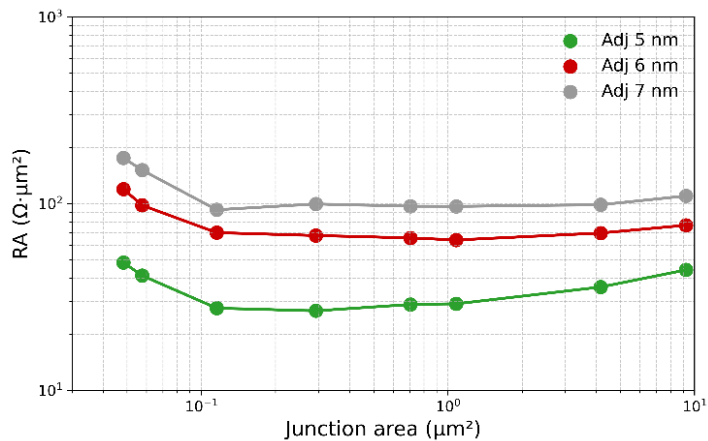

FIGURE 6.  RA product using measured resistance values, with adjustment for lead resistances versus nominal junction areas.

It can be seen from the location of voltage taps in the layout images shown in Fig. 2(a,b) that the 4-wire Kelvin resistance measurement still includes some extraneous resistance at room temperature that would be zero below the superconducting transition temperature of α–Ta. Because α–Ta has a room-temperature resistivity an order of magnitude higher than aluminum, even such small additions from these leads significantly contribute to the total resistance measured [20]. By subtracting the resistance measured on another wafer that has no ALD TaN tunnel barrier, it is possible to estimate the resistance contribution from the junction itself. This estimate of the junction resistance can be multipled by the nominal junction area to determine the room temperature resistance-area (RA) product. The junction area dependance of the median estimated RA product at room temperature is shown in Fig. 6 for junctions using 5, 6 and 7 nm ALD TaN barriers.

The RA product shows a sharp increase at smaller junction areas. The room temperature RA product shown in Fig. 6 is influenced by a few factors. Firstly, the increase of RA for junctions with area < 0.05 $\mu m^2$ is affected by deviations from the design value of junction area, since the reticle fabrication did not include optical proximity corrections (OPC). Additionally, junction sizes were estimated by applying a fixed value of process-induced dimensional bias determined from metrology structures, rather than measuring the junctions themselves. Secondly, as shown in Fig. 2(a), our design has voltage taps a few microns away from the junction, contributing to lead resistance. Such layout considerations contribute to higher lead resistance effects for Ta vis-à-vis Al based wiring, due to its order of magnitude higher room temperature resistivity. Thirdly, variations in Ta thickness from wafer to wafer due to CMP induced dishing can contribute to uncompensated lead resistances that affect wider Ta lines to a greater extent [20]. Despite these effects, it can be seen that the RA product varies from ~25 to 100 $\Omega \cdot \mu m^2$ as the thickness increases from 5 nm to 7 nm. These values are comparable to those reported in the literature for qubit applications [33].

The room temperature and cryogenic characterization by broad-band spectroscopic ellipsometry of the ALD TaN films on thermally grown silicon oxide confirm it to be an insulator with no signature of free-carrier absorption even at room temperature [26]. Hence, conduction through the ALD TaN happens through tunneling both at room temperature and at cryogenic temperatures.

## B. Cryogenic Characterization

I-V characteristics of junctions of varying dimensions (using 4 nm ALD TaN) are shown in Fig. 7(a), measured by sourcing a current and measuring the resulting voltage across the junction in a 4-wire Kelvin configuration. A small (0.06 to 1.04 mV) voltage offset was measured for junctions in the superconducting state below their critical current. This was also seen in superconducting nanowires measured with the same hardware and hence is assigned to issues related to measurement electronics and has been subtracted in Fig. 7(a). After the subtraction, the highest measured gap voltage is 0.745 mV, 60% of the 1.25 mV value predicted for junctions using α-Ta electrodes. The gap voltage decreases towards 0.7 mV for larger junctions, with the decrease attributed to self-heating that is increasingly an issue as junction size increases.

Similar to the analysis by Tolpygo et al. [13] for aluminum oxide-based junctions, extrapolation of the normal regime to the y-axis goes through the zero-current point, indicating no extraneous current mechanisms are present that give rise to 'excess current'.

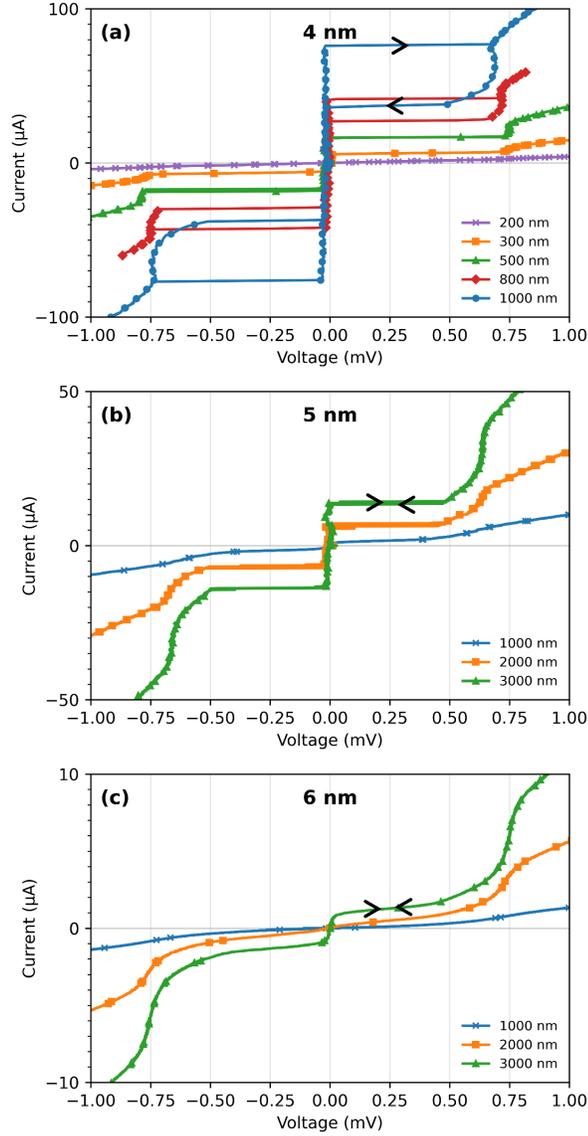

FIGURE 7. I-V characteristics of Josephson junctions using (a) 4 nm ALD TaN, (b) 5 nm ALD TaN, and (c) 6 nm ALD TaN in the tunnel barriers. The legend values (200 nm, 300 nm, 1000 nm, etc.) correspond to the nominal lateral dimensions of the square junctions. Black arrows indicate the current sweep direction, showing the transition from the superconducting to normal state and the reverse on the return branch.

It can be seen in Fig. 7(a) that hysteresis is present in the I–V traces of junctions larger than 500 nm (nominal). The critical current ($I_c$) is defined as the maximum supercurrent the junction can carry before switching to the resistive state, while the retrapping current ($I_r$) is the current at which the junction returns to the superconducting state upon decreasing bias. The separation between the critical current and the retrapping current increases with junction size. This is indicative of enhanced self-heating, as larger junctions support higher critical currents and therefore more dissipation due to joule heating effect [27]. When junctions using 5 nm ALD TaN are considered, as shown in Fig. 7(b), hysteresis is not seen even at the larger dimensions tested – attributable to the lower critical current density and lower self-heating in junctions with a thicker barrier.

In Fig. 7(c) (as in Fig. 7(a)), the voltage axis has been shifted to remove the measurement-related offset observed at zero current. For junctions with the thicker 6 nm ALD TaN barrier, the reduced critical currents approach the noise floor and resolution limits of source–measure instrumentation. As a result, junctions smaller than 1 μm could not be reliably characterized, and even the 1 μm junction data in Fig. 7(c) are near the limits of measurement accuracy. Only the 3 μm x 3 μm junctions show a characteristic from which a critical current can be reliably extracted with the voltage axis DC-shifted to zero, as described previously. The gap voltage is about 0.75 mV, higher than for corresponding junctions with thinner ALD TaN, attributed to lower self-heating effects. By choosing the point where the dI/dV vs $I$ curve first reaches a low value, a critical current of 1.0 μA was determined.

The variation of critical current with (nominal) junction area is shown in Fig. 8. Panel (a) of this figure indicates that for JJs using 4 nm ALD TaN, the critical current density is 76 μA/μm², while panel (b) shows that for junctions with 5 nm ALD TaN, the critical current density ($J_c$) is 1.53 μA/μm².

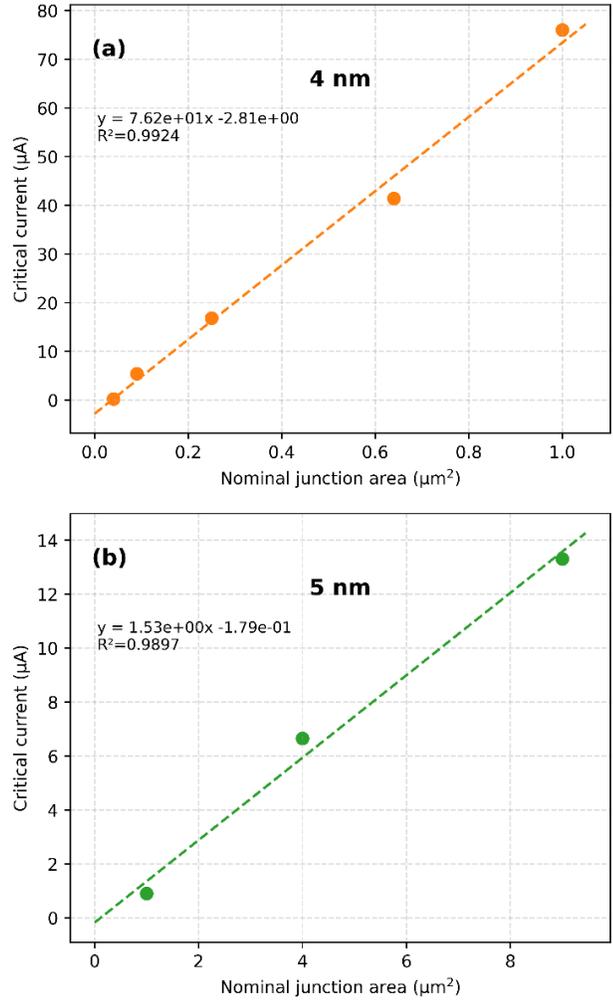

FIGURE 8. Critical current as a function of nominal junction area using (a) 4 nm ALD TaN and (b) 5 nm ALD TaN in the tunnel barrier.

Fig. 9 shows the I-V characteristics of a junction with 4 nm ALD TaN as a function of temperature (from 100 mK to 3.8 K). The $I_c$ of the junction stays relatively unchanged from mK temperatures to about 1.5 K. The stability of $I_c$ over this range allows measurements conducted in the ADR to be used to predict performance when the junctions are operated at higher temperatures, as might be the case for SFQ circuits used to control qubit chips [6]. The temperature insensitivity of $I_c$ at low temperatures is consistent with the Ambegaokar–Baratoff (A-B) relation [28,29] for SIS, which predicts

$$I_c R_n = \frac{\pi \Delta(T)}{2e} \tanh\left(\frac{\Delta(T)}{2k_B T}\right),$$

reducing to $I_c R_N = \pi\Delta(0)/2e$ at T = 0 K.

The value of experimentally measured $eI_c R_N/2\pi k_B T_c$ is slightly lower than that expected from A-B theory. This deviation can be explained from the fact that the junctions experienced self-heating effects due to weak thermal conductance between the device chip and the sample package as described earlier. Due to these heating effects, the observed $I_c$ is lower than the actual $I_c$ The reported value of the temperature (Tmeasured) comes from a measurement near the cold-finger, and is likely lower than the actual temperature of the junction – with the deviation being greater as temperature rises.

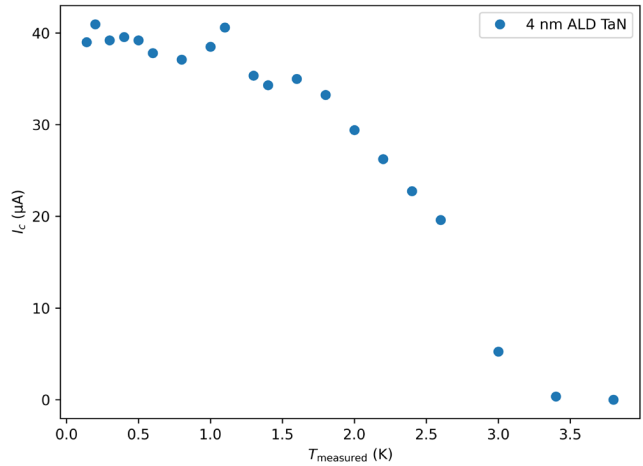

FIGURE 9.     Temperature dependance of the critical current for an  800 nm x 800 nm (nominal) Josephson junction with a 4 nm ALD TaN tunnel barrier.

Fig. 10 shows the I-V traces measured on junctions using a 2 nm thick (nominal) ALD TaN film. The I-V trace of a 500 nm square junction is affected by the design of the structure, where lead wires switches to normal state before the junction itself, as shown in Fig. 10(a). This results in the true gap voltage being obscured. For a 300 nm square junction, the junction critical current is small enough that a gap voltage consistent with theoretical limits is observed [30]. The gap voltages are much lower than for junctions with thicker ALD TaN, indicative of self-heating, due to poor thermal coupling through the package. In Fig. 10 (b), I-V characteristics of rectangular junctions (1 μm x 120 nm) are shown, with physically reasonable gap voltages. This is attributed to the leads being sufficiently wide, and hence not limiting the critical current density, as can be seen in the inset figure. Hence such thin ALD TaN films, with estimated current densities in the range of 350 - 600 μA/μm$^2$ need to be fabricated anew, with improved layouts, along with packaging designed to provide better thermal coupling to the cold stage.

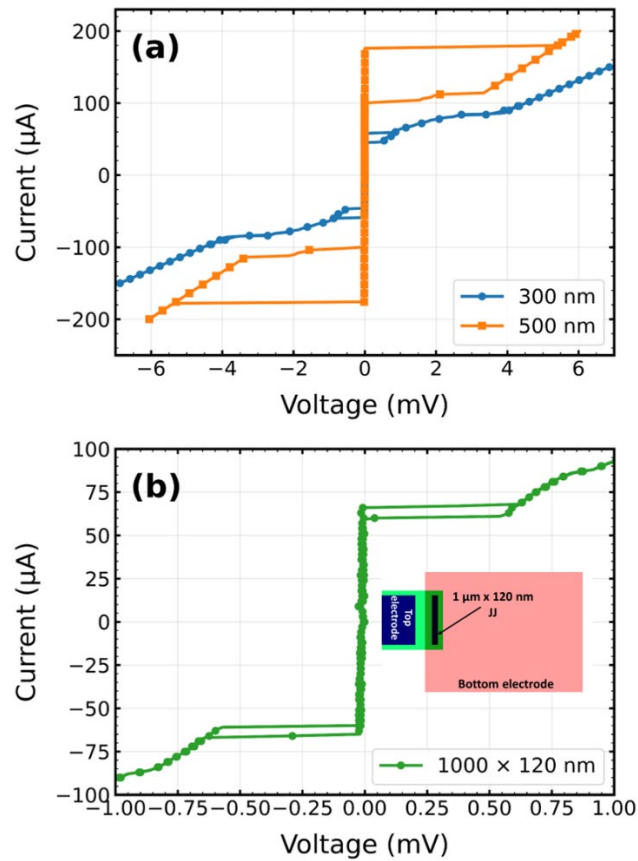

FIGURE 10. I-V characteristics of junctions using 2 nm ALD TaN, with (a) showing square junctions, and (b) showing results from rectangular junctions aligned to have widest lead wires.

In Fig. 11, $J_c$ as a function of thickness is presented for 4, 5, and 6 nm ALD TaN based junctions. For all junctions, $J_c$ is determined by adding a fixed process induced dimensional bias of 40 nm to the nominal junction.

The critical current density decreases exponentially with ALD TaN thickness, as expected for tunneling. Fitting $J_c(t) = J_0 e^{-\beta t}$, (excluding 2 nm, which shows self-heating) gives $\beta = 3.28$ nm$^{-1}$, corresponding to an attenuation length $\lambda = 0.305$ nm ($\lambda = 1/\beta$). The fit confirms the expected exponential trend for Josephson coupling [31].

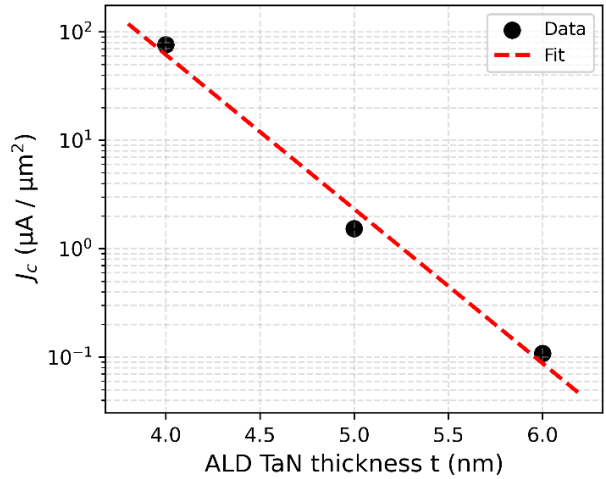

FIGURE 11. Critical current density as a function of ALD TaN thickness on a log-linear plot.

A set of junctions that had been previously characterized was retested, after storage at room temperature for 4 months in a container with ambient air pressure. In Fig. 12, I-V traces taken 140 days apart are shown for 300 nm and 1 μm junctions, using a 4 nm ALD TaN. The lack of a significant difference between the two measurements indicates indicates weak aging exhibited by junctions using ALD TaN as the tunnel barrier. It is noted that the first measurement was done 10 weeks after JJ formation, due to time required for dicing, packaging, and scheduling test time.

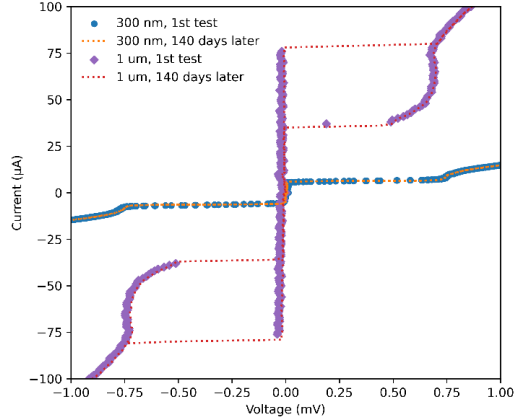

FIGURE 12.   I-V characteristics showing lack of aging over 140 days for two junctions using 4 nm ALD TaN.

For comparison, Al/AlO$_x$/Al junctions fabricated without special substrate cleaning have been reported to double or triple in resistance within weeks under ambient conditions, whereas even with optimized cleaning procedures they remain stable for ~20 weeks [32]. Also, Al/AlO$_x$/Al junctions fabricated using 300 mm processes exhibited 3.7 % change in resistance over 146 days (~21 weeks) under clean-room storage [24]. In contrast, our α-Ta/ALD TaN junctions show no measurable degradation over ~20 weeks of ambient storage, even without any pre-measurement cleaning, which highlights the stability of the ALD TaN barrier.

## IV. CONCLUSION

We have demonstrated, for the first time, the fabrication of Josephson junctions using superconducting α-Ta electrodes and an insulating ALD TaN tunnel barrier on 300 mm wafers, utilizing a CMOS-compatible process flow. This work is the first step towards a scalable and thermally robust junction technology that can be adapted for both superconducting qubits and single-flux-quantum (SFQ) digital circuits.

Room-temperature and cryogenic electrical measurements confirm yields exceeding 95% for 200 nm square junctions at multiple barrier thicknesses. For 4 nm ALD TaN barriers, a critical current density of 76 µA/µm² was achieved, and a measured gap voltage ~60% of the theoretical maximum. Thickness scaling from 4 to 6 nm ALD TaN follows the expected exponential dependence and therefore confirms the tunneling behavior of these JJs.

Devices with thin (2 nm ALD TaN) barriers have I-V characteristics affected by self-heating effects due to much higher $I_c$ values and require improved thermal coupling and optimized packaging to mitigate self-heating and allow accurate measurement. Long-term stability testing over 140 days reveals negligible aging effects, underscoring the robustness of the ALD TaN barrier.

This work to develop Josephson junctions fabricated at 300 mm wafer scale, using α-Ta electrodes and ALD TaN in the tunnel barrier is the starting point of ongoing efforts to create a non-profit 300 mm Quantum Foundry. Such a Foundry will help advance the field by providing researchers and designers in academia, national laboratories and start-up companies with a stable, well-characterized process associated with a Superconducting Process Design Kit (SQPDK).  Utilizing the SQPDK, novel quantum circuits could be designed with predictable circuit performance after fabrication, such that the multi-project wafer (MPW) and fabless semiconductor paradigms that revolutionized microelectronics could be used to create a similar flourishing of superconducting electronics.

Ongoing work is focused on reducing self-heating in devices through improved packaging, quantifying sub-gap performance, and testing these JJs in qubit and SFQ circuits.

This technology provides a path toward a scalable, wafer-scale manufacturing platform for superconducting quantum and digital circuits.

# ACKNOWLEDGMENT

This work was funded in part by Air Force Research Laboratory (through contract numbers FA8750-19-1-0031, FA8750-20-P-1710, and FA864921P0773), and by NY Creates. The authors are thankful to Peter McGarvey at the Rochester Test, Assembly and Packaging facility for help with dicing, and to Andrew Brownell and Nick Barton at AFRL Rome for wire-bonding and packaging the devices for cryogenic testing. Discussions with Igor Vernik and Oleg Mukhanov on self-heating effects on measured I-V characteristics are gratefully acknowledged. Approved for Public Release; Distribution Unlimited; Case Number: AFRL-2025-5168. Any opinions, findings, and conclusions or recommendations expressed in this material are those of the author(s) and do not necessarily reflect the views of Air Force Research Laboratory.

# REFERENCES

[1] M. Devoret and R. Schoelkopf, "Superconducting Circuits for Quantum Information: An Outlook," Science, vol. 339, pp. 1169, 2013, doi: 10.1126/science.1231930.
[2] Q. Herr, T. Josephsen, and A. Herr, "Superconducting pulse conserving logic and Josephson-SRAM," Appl. Phys. Lett., vol. 122, pp. 182604, 2023, doi: https://doi.org/10.1063/5.0148235.
[3] C. A. Hamilton et al., "Operation of NIST Josephson array voltage standards," Journal of Research of the National Institute of Standards and Technology, vol. 95, no. 3, pp. 219, 1990, doi: 10.6028/jres.095.026.
[4] A. W. Kleinsasser et al., "Dependence of critical current density on oxygen exposure in Nb-AlO$_x$-Nb tunnel junctions," IEEE Transactions on Applied Superconductivity, vol. 5, no. 1, pp. 26-30, March 1995, doi: 10.1109/77.384565.
[5] J. H. Han et al., "Growth of tantalum nitride film as a Cu diffusion barrier by plasma-enhanced atomic layer deposition from bis((2- (dimethylamino)ethyl)(methyl) amido) methyl(tertbutylimido)tantalum complex," Appl. Surf. Sci., vol. 362, pp. 176–181, Jan. 2016, doi: 10.1016/j.apsusc.2015.11.095.
[6] R. McDermott et al., "Quantum–classical interface based on single flux quantum digital logic," Quantum Sci. Technol., vol. 3, no. 2, pp. 024004, 2018, doi: 10.1088/2058-9565/aaa3a0.

[7] C. Wang et al., "Towards practical quantum computers: transmon qubit with a lifetime approaching 0.5 milliseconds," npj Quantum Inf, vol. 8, no. 3, 2022, doi: 10.1038/s41534-021-00510-2.

[8] M. P. Bland et al. "2D transmons with lifetimes and coherence times exceeding 1 millisecond." arXiv preprint arXiv:2503.14798 (2025).

[9] A.P.M. Place et al., "New material platform for superconducting transmon qubits with coherence times exceeding 0.3 milliseconds. Nat Commun, vol. 12, pp. 1779, 2021. doi: 0.1038/s41467-021-22030-5.

[10] D. Edelstein et al., "Full copper wiring in a sub-0.25 /spl mu/m CMOS ULSI technology," Internat. Electron Dev. Meeting Technical Digest, Washington, DC, USA, pp. 773-776, 1997, doi: 10.1109/IEDM.1997.650496.

[11] D. Dochev et al., "The influence of aging and annealing on the properties of Nb/Al-AlO$_x$/Nb tunnel junctions," Journal of Physics: Conference Series, vol. 234, no. 4, p. 042006, 2010. doi: 10.1088/1742-6596/234/4/042006.

[12] N. D. Korshakov et al., "Aluminum Josephson junction microstructure and electrical properties modified by thermal annealing." Scientific Reports vol. 14, no. 1, pp. 26066 2024, doi: 10.1038/s41598-024-74071-7.

[13] S. K. Tolpygo et al., "Advanced Fabrication Processes for Superconducting Very Large-Scale Integrated Circuits," IEEE Transactions on Applied Superconductivity, vol. 26, no. 3, pp. 1-10, April 2016, Art no. 1100110, doi: 10.1109/TASC.2016.2519388.

[14] J. Wilt et al., "Atomically thin Al 2 O 3 films for tunnel junctions." Physical Review Applied vol. 7, no. 6, pp. 064022, 2017. doi: 10.1103/PhysRevApplied.7.064022.

[15] M. A. Wolak, Matthaeus et al., "SNS Josephson junctions with tunable Ta–N barriers," IEEE Transactions on Applied Superconductivity vol. 29, no. 5, pp. 1-4, Aug. 2019, Art no. 1102204, doi: 10.1109/TASC.2019.2904489.

[16] H. Kim, C. Detavenier, O. Van Der Straten, S. M. Rossnagel, A. J. Kellock, and D-G. Park. "Robust TaNx diffusion barrier for Cu-interconnect technology with subnanometer thickness by metal-organic plasma-enhanced atomic layer deposition." Journal of Applied Physics 98, pp. 014308 (2005), doi: 10.1063/1.1935761.

[17] Li Yongliang and Xu Qiuxia. "TaN wet etch for application in dual-metal-gate integration technology." J. Semicond. vol.30, no. 12, pp. 126001 2009, doi: 10.1088/1674-4926/30/12/126001.

[18] Z. Tang et al., "Impact of TaN as Wet Etch Stop Layer on Device Characteristics for Dual-Metal HKMG Last Integration CMOSFETs," in IEEE Electron Device Letters, vol. 34, no. 12, pp. 1488-1490, Dec. 2013, doi: 10.1109/LED.2013.2287271.

[19] Likharev, K.K. (1986). Dynamics of Josephson Junctions and Circuits (1st ed.). Routledge. doi: 10.1201/9781315141572

[20] E. Bhatia et al., "Chemical mechanical planarization for Ta-based superconducting quantum devices." J. Vac. Sci. Technol. B vol. 41, no. 3, pp. 033202, 2023, doi: 10.1116/6.0002586.

[21] H. Frost et al., "Adjusting tunnel junction characteristics of atomic layer deposition films through underlayer modification," PCT Application WO2025006731, filed June 27,2024. https://patentscope.wipo.int/search/en/detail.jsf?docId=WO2025006731.

[22] J. Verjauw et al., "Path toward manufacturable superconducting qubits with relaxation times exceeding 0.1 ms," npj Quantum Information, vol. 8, no. 1, pp. 93 2022, doi: 10.1038/s41534-022-00600-9.

[23] M. Osman, et al., "Fabrication and characterization of reproducible Al/AlOx/Al Josephson junctions for superconducting quantum circuits," Appl. Phys. Lett., vol. 118, pp. 064002, 2021, doi: 10.1063/5.0037093

[24] J. Van Damme et al., "Advanced CMOS manufacturing of superconducting qubits on 300 mm wafers," Nature, vol. 634, pp. 74–79, 2024. doi: 10.1038/s41586-024-07941-9.

[25] E. Bhatia et al., "Enabling 300 mm Wafer-Scale Fabrication of Superconducting Quantum Devices," 35th Annual SEMI Advanced Semiconductor Manufacturing Conference (ASMC) (IEEE, 2024), pp. 1–6, 2024, doi: 10.1109/ASMC61125.2024.10545509.

[26] A. L. Gonzalez et al., "Dielectric Function of Tantalum Nitride Formed by Atomic Layer Deposition on 300 mm Wafers for Josephson junction applications." APS March Meeting Abstracts, vol. 2024, pp. K15-003. 2024.

[27] Oikawa, K., Y. Kamiya, and M. Toyota. "Numerical analysis of temperature and current distributions in large-size intrinsic Josephson junctions with self-heating." AIP Advances vol. 10, pp. 085113, 2020, doi: 10.1063/5.0018989.

[28] V. Ambegaokar and A. Baratoff, "Tunneling between superconductors," Phys. Rev. Lett., vol. 10, no. 11, pp. 486–489, 1963, doi: 10.1103/PhysRevLett.10.486.

[29] V. Ambegaokar and A. Baratoff, "Tunneling between superconductors" (Erratum), Phys. Rev. Lett., vol. 11, no. 3, p. 104, 1963, doi: 10.1103/PhysRevLett.11.104.

[30] D.R Tilley and J. Tilley, *Superfluidity and Superconductivity* (Adam Hilger Ltd., Bristol), 1990, 10.1201/9780203737897

[31] M. Tinkham, *Introduction to Superconductivity*, 2nd ed. (McGraw-Hill, New York, 1996).

[32] I. M. Pop et al., "Fabrication of stable and reproducible submicron tunnel junctions," J. Vac. Sci. Technol. B vol. 30, pp. 010607, 2012, doi: 10.1116/1.3673790.

[33] Y. Nakamura et al., "Superconducting qubits consisting of epitaxially grown NbN/AlN/NbN Josephson junctions", Appl. Phys. Lett. vol. 99, pp 212502, 2011, doi:10.1063/1.3663539.